\documentclass{article}
\usepackage{graphicx}
\def\err#1#2{\lower2pt\hbox{$\stackrel{\scriptstyle +#1}{\scriptstyle -#2}$}}

\begin{document}
\centerline{\today \hfill{UCD-HEP-99-15}}
\begin{center}
{\bf \large $B_c$ meson production at the Tevatron Revisited} \\
Kingman Cheung \\
{\it Department of Physics, University of California, Davis, CA 95616 USA}
\end{center}

\begin{abstract}
CDF recently measured the quantity $\frac{\sigma(B_c^+)}{\sigma(B^+)}
\frac{{\rm BR}(B_c^+ \to J/\psi \ell^+ \nu)}{{\rm BR}(B^+ \to J/\psi K^+)}$, 
{}from which we determine the ratio $\frac{\sigma(B_c^+)}{\sigma(\bar b)}$ 
to be $(2.08\err{1.06}{0.95})\times 10^{-3}$.  In this note, we show that 
the ratio $\frac{\sigma(B_c^+)}{\sigma(\bar b)}$ obtained by dividing the 
$\sigma(B_c^+)$ by the leading order $\sigma(\bar b)$ is consistent with 
this derived CDF measurement.  We calculate the cross section $\sigma(B_c^+)$ 
using the perturbative  QCD fragmentation functions of 
Braaten, Cheung, and Yuan and the corresponding induced gluon fragmentation 
functions,  with the charm-quark mass $m_c$ as a parameter.   We also 
estimate the parameter $m_c$ from the CDF data and then predict the 
production rate at RunII.
\end{abstract}

\section{}
CDF \cite{cdf-bc} and LEP Collaborations \cite{lep-bc} recently published 
their results in search for the  final heavy-heavy quark bound state
-- charmed-beauty meson ($B_c$).  The most impressive is the result by CDF,
which established a signal of $4.8 \sigma$ (from a null hypothesis), using 
the semi-leptonic decay channels of the $B_c$ meson, $B_c^+ \to J/\psi \ell^+
\nu$, with $\ell=e,\mu$.  CDF measured the ratio
\begin{equation}
\label{data}
{\cal R} \equiv
\frac{\sigma(B_c^+)}{\sigma(B^+)}
\frac{{\rm BR}(B_c^+ \to J/\psi \ell^+ \nu)}{{\rm BR}(B^+ \to J/\psi K^+)}
= 0.132 \err{0.041}{0.037}\, ({\rm stat.}) \; \pm 0.031\,({\rm syst.})
\; \err{0.032}{0.020}\,({\rm lifetime}) \;,
\end{equation}
where in ${\rm BR}(B_c^+ \to J/\psi \ell^+ \nu)$ the branching ratios for
$e$ and $\mu$ are assumed equal, and 
the last error comes from the error in the measurement of the $B_c$ lifetime.
Based on the following data from Particle Data Book \cite{pdg}
\begin{equation}
\label{2}
{\rm BR}(B^+ \to J/\psi K^+) = (9.9\pm1.0)\times 10^{-4}\;,\;\;
\frac{\sigma(B^+)}{\sigma(\bar b)} =0.397 \err{0.018}{0.022} \;,
\end{equation}
and a theoretical calculation \cite{bc-decay}
\begin{equation}
{\rm BR}(B_c^+ \to J/\psi \ell^+ \nu) = 2.5\pm0.5 \%
\end{equation}
we are able to deduce this ratio 
\begin{equation}
\label{bcfrag}
\frac{\sigma(B_c^+)}{\sigma(\bar b)} = \frac{ {\rm BR}(B^+ \to J/\psi K^+)}
{{\rm BR}(B_c^+ \to J/\psi \ell^+ \nu)}\;
\frac{\sigma(B^+)}{\sigma(\bar b)} \times  {\cal R} = 
\left( 2.08 \err{1.06}{0.95}\right )\times 10^{-3} \;,
\end{equation}
where the error is obtained by adding the relative errors in quadrature.
Note that 
the ratio $\frac{\sigma(B^+)}{\sigma(\bar b)}$ in Eq. (\ref{2}) quoted in 
Particle Data Book represents the fraction of $\bar b$ that hadronizes 
into a $B^+$ meson, which was measured at LEP.  This fraction is, to a 
good approximation, independent of $p_T$ cuts.
Thus the ratio $\frac{\sigma(B_c^+)}{\sigma(\bar b)}$ that we obtained 
in Eq. (\ref{bcfrag}) represents the ratio of the cross section of $B_c^+$ to 
the cross section of $\bar b$ under the same $p_T$ cut as the CDF measurement
${\cal R}$.
This ratio has no direct implication that $B_c^+$ meson
is produced directly from $\bar b$, which is in contrast to $B^+$ meson
that $B^+$ meson is, in general, assumed coming from the fragmentation of 
$\bar b$.

The purpose of this note is to verify that the ratio in Eq. (\ref{bcfrag}) 
is consistent with $\sigma(B_c^+)$ calculated using the perturbation QCD 
fragmentation functions for $\bar b \to B_c^+$ 
\cite{s-wave,bcfrags_others,p-wave,ours} and the corresponding induced gluon
fragmentation functions, as well  as using the leading order (LO) b-quark
production.  We shall also obtain the range of the parameters involved
in the fragmentation functions.  Once we obtain the parameters
we can then predict the production rate for the RunII at
the Tevatron, where much higher statistics can be accumulated.

The fragmentation approach employed here is different from the
full tree-level $\alpha_s^4$ calculation \cite{had}.  
%
%
Comparison between these two approaches were made in some of the papers
in Refs. \cite{had}. Basically, the fragmentation approach gives a reasonable
approximation to the full tree-level calculation as long as $p_T > 2 M_{B_c}$.
Nevertheless, one disadvantage of the full calculation is that higher order 
effects cannot be easily included unless the NLO calculation is performed.  
On the other hand,
using the fragmentation approach some important higher order effects can be
included, namely, the contribution from gluon fragmentation and the
contribution from higher orbital states below the BD threshold.  
We shall show that including these contributions
we can easily account for the ratio in Eq. (\ref{bcfrag}) without 
employing extreme parameters.

\section{}
In this section, we remind the readers about the importance of the 
induced gluon fragmentation.  The gluon fragmentation function for
$g \to B_c^+$ at the initial scale (heavy quark scale) is $O(\alpha_s)$ 
smaller than the heavy quark fragmentation function for $\bar b\to B_c^+$.  
Thus, the main source of gluon
fragmentation comes from the Altarelli-Parisi evolution of the heavy quark
fragmentation function.  
The Altarelli-Parisi evolution equations for the fragmentation functions are
\begin{eqnarray}
\label{AP1}
\mu \frac{\partial}{\partial \mu} D_{\bar b\to H}(z,\mu) &=& 
\int_z^1 \frac{dy}{y}
P_{\bar b\to \bar b}(z/y,\mu)\; D_{\bar b \to H}(y,\mu) + 
\int_z^1 \frac{dy}{y} P_{\bar b\to g}(z/y,\mu)\; D_{g \to H}(y,\mu)  \\
\label{AP2}
\mu \frac{\partial}{\partial \mu} D_{g\to H}(z,\mu) &=& \int_z^1 \frac{dy}{y}
P_{g \to \bar b}(z/y,\mu)\; D_{\bar b \to H}(y,\mu) +
\int_z^1 \frac{dy}{y} P_{g \to g}(z/y,\mu)\; D_{g \to H}(y,\mu)
\end{eqnarray}
where $H$ denotes any $(\bar b c)$ states, and $P_{i\to j}$ are the usual
Altarelli-Parisi splitting functions. 
%
The initial scale heavy quark and gluon fragmentation functions are 
\cite{s-wave}
\begin{eqnarray}
\label{DBc} 
D_{{\bar b} \rightarrow \bar b c(n^1S_0)}(z,\mu_0)
& = & {2 \alpha_s(2 m_c)^2 |R_{nS}(0)|^2 \over 81 \pi m_c^3} 
\; {r z (1-z)^2 \over (1 - \bar r z)^6} \;\;
\Bigg[ 6 \;-\; 18(1-2r)z \nonumber \\
&+& \;(21-74r+68r^2)z^2
- \, 2\bar r(6-19r+18r^2)z^3 \, + \, 3\bar r^2(1-2r+2r^2)z^4 \Bigg] 
\;, \\
\label{DBst}
D_{{\bar b} \rightarrow \bar b c(n^3S_1)}(z,\mu_0)
& =& {2 \alpha_s(2m_c)^2 |R_{nS}(0)|^2 \over 27 \pi m_c^3} \; 
{r z (1-z)^2 \over (1 - \bar r z)^6} \;\;
\Bigg[ 2 \;-\; 2(3-2r)z \nonumber \\
&+& 3(3-2r+4r^2)z^2  \;-\; 2\bar r(4-r+2r^2)z^3
+ \bar r^2(3-2r+2r^2)z^4 \Bigg] \;, \\
D_{g\to B_c}(z,\mu) &=& D_{g\to B_c^*}(z,\mu) = 0 \;\;\;\;\; {\rm for}
\;\;\;\;\;\;  \mu \le 2(m_b+m_c) \;,
\end{eqnarray}
where $r=m_c/(m_b+m_c)$, $\bar r=1-r$, $\mu_0=m_b+2m_c$, and 
$R(0)$ is the radial wavefunction at the origin. They are the initial
boundary conditions to the evolution equations in 
Eqs. (\ref{AP1})--(\ref{AP2}).
Here we only give the S-wave fragmentation functions, which contribute
dominantly to $B_c$ production, the P-wave fragmentation functions can be
found in Ref. \cite{p-wave}.  Nevertheless, P-wave fragmentation functions
contribute only at 10\% level to the total $B_c$ production.
The less determined parameters in the above functions are $|R(0)|$ and
$m_c$.  The value for $R(0)$ can be determined in a potential-model calculation
\cite{wave}.  In Ref. \cite{wave}, $|R_{1S}(0)|^2$ ranges from 1.5 to 1.7 
GeV$^3$ (the extreme value of 3.2 GeV$^3$ is not used here.)
The fixed input parameters of our present calculation are tabulated in Table
\ref{table1}, while $m_c$ is chosen as a variable parameter in our calculation,
because the fragmentation function is very sensitive to $m_c$, which appears
as $m_c^3$ in the denominator: see Eqs. (\ref{DBc})--(\ref{DBst}).  
Overall, we include all $n=1$ S-wave and P-wave, and $n=2$ S-wave 
states, which are below the BD threshold \cite{wave}, in our calculation.

\begin{table}[th]
\centering
\begin{tabular}{|c|cc|}
\hline
       &   $n=1$    &  $n=2$ \\
\hline
\hline
$m_b$    &  4.9 GeV  &  4.9 GeV \\
$R_{nS}(0)$   &  1.28 GeV$^{3/2}$  &   0.99 GeV$^{3/2}$ \\
$H_1$      &  10 MeV   &   - \\
$H_8'(m)$     &  1.3 MeV   &   - \\
$\cos\theta_{\rm mix}$  & 0.999 & -  \\
\hline
\end{tabular}
\caption{\small Input parameters to the perturbative QCD fragmentation 
functions for $n=1$ and $n=2$.  $H_1,H'_8(m),\cos\theta_{\rm mix}$ are 
parameters for P-wave states, see Ref. \cite{p-wave}.
\label{table1} }
\end{table}

\section{}
We are ready to compute the ratio $\sigma(B_c^+)/\sigma(\bar b)$ with
$\sigma(B_c^+)$ calculated by the fragmentation approach and $\sigma(\bar b)$
by the LO calculation.  The 
``improved'' tree-level cross section for $B_c^+$ is given by
\begin{equation}
\sigma(B_c^+) = \sum_{ij}\;
\int dx_1 dx_2 dz \; f_i(x_1) f_j(x_2) \, \Biggr[
\hat \sigma(ij \to \bar b X,\mu) \; D_{\bar b \to B_c^+}(z,\mu) +
\hat \sigma(ij \to g X,\mu) \; D_{ g \to B_c^+}(z,\mu)  \Biggr ] \;,
\end{equation}
where $\mu$ is the factorization scale and is chosen to be
$\mu=\mu_T \equiv \sqrt{p_{T_{\bar b}}^2 + m_b^2}$.
We called this the improved cross section because it includes higher
order corrections from gluon fragmentation.  For $\bar b$ cross section we
use the LO calculation.  When we calculate the ratio of cross sections,
the dependence on factorization scale, higher-order QCD corrections, 
parton distribution functions, and
$m_b$ are substantially reduced.  We anticipate the ratio $\sigma(B_c^+)/
\sigma(\bar b)$ calculated at tree-level is reasonably accurate without
a NLO calculation, providing that the present error on $B_c$ production 
is very large.  

\begin{figure}[th]
\centering
\includegraphics[width=6in]{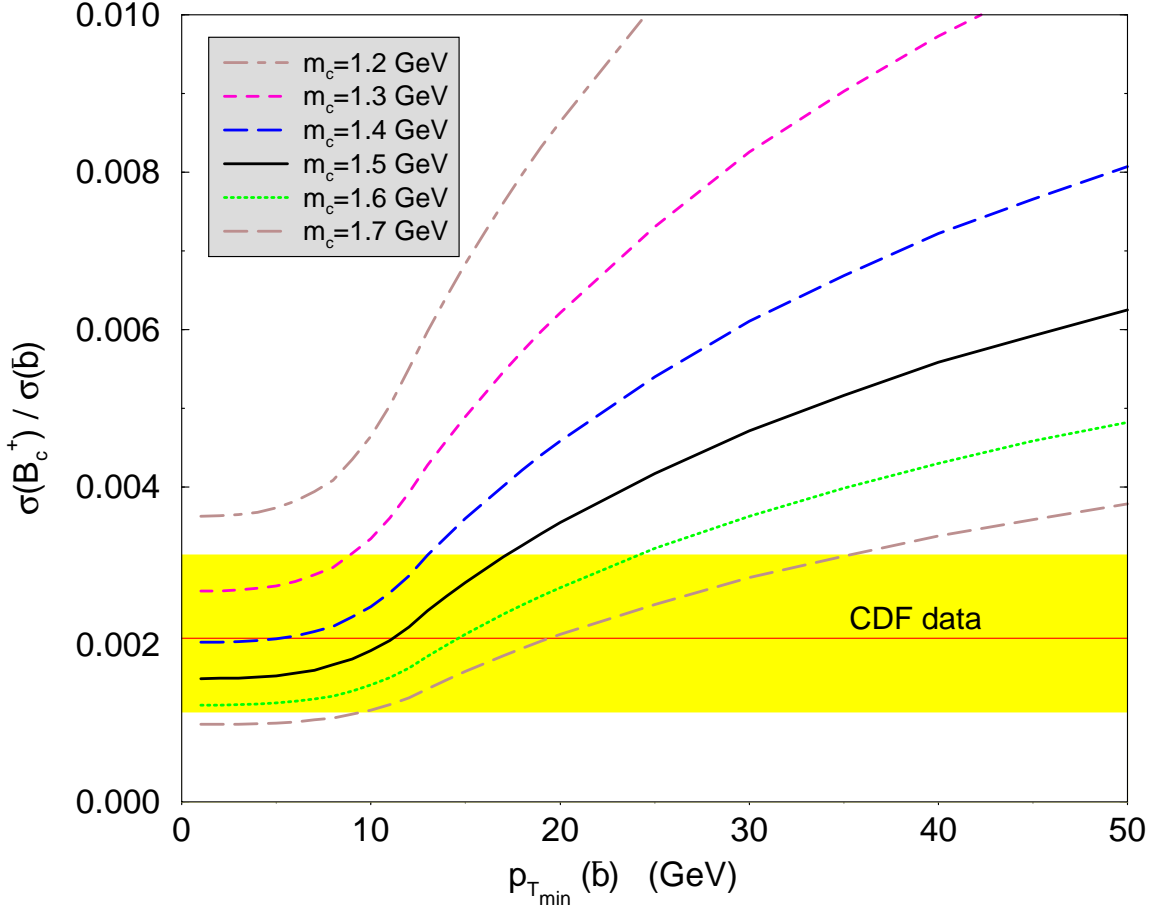}
\caption{\small 
The ratio of $\sigma(B_c^+)/\sigma(\bar b)$ versus $p_{T_{\rm min}}$ cut on
$\bar b$, calculated by fragmentation approach at the Tevatron:
$\sqrt{s}=1.8$ TeV.  A rapidity cut of $|y|<1$ is imposed. 
The shaded band is the data in Eq. (\ref{bcfrag}), which is derived from 
the CDF data in Eq. (\ref{data}).
\label{fig1} 
}
\end{figure}

We show the ratio $\sigma(B_c^+)/\sigma(\bar b)$ versus 
$p_{T_{\rm min}}(\bar b)$ for $m_c=1.2 - 1.7$ GeV in Fig. \ref{fig1}.  We
note that this ratio increases with $p_{T_{\rm min}}(\bar b)$, due to 
the induced gluon fragmentation contribution.  
When $p_{T_{\rm min}}(\bar b)$
increases, the scale of the fragmentation function rises and, therefore,
the induced gluon fragmentation function also increases.  
If we did not include the induced gluon fragmentation contribution, the 
ratio $\sigma(B_c^+)/\sigma(\bar b)$ would have been a constant, 
giving rise to a horizontal line coincide with the lower part of the 
corresponding curve in Fig. \ref{fig1}.   
Although the gluon fragmentation
probability is much smaller than the $\bar b$ fragmentation, the production
by gluon fragmentation turns out not negligible, because the amplitude
squared of the most important subprocess $gg\to gg$ is more than an order
of magnitude larger than that of $gg \to b \bar b$ \cite{ours}.
Figure \ref{fig1} also shows the sensitivity to $m_c$.

We put the band of $\sigma(B_c^+)/\sigma(\bar b)$ given by Eq. (\ref{bcfrag})
onto Fig. \ref{fig1}.
We note that the CDF data in Eq. (\ref{data}) is for $B_c^+$ and $B^+$ with
$p_T> 6.0$ GeV and $|y|<1$ \cite{cdf-bc}.  We have to convert this $p_T$
requirement on $B^+$ and $B_c^+$ to $p_T$ requirement on $\bar b$, because 
the fragmentation spectrum of $\bar b$ is not monochromatic.  The average
momentum fraction $\langle z \rangle$ for fragmentation of $\bar b$ into
$B^+$ and $B_c^+$ is about $0.7-0.8$ at the scale $\mu \approx 8-10$ GeV.
Hence, the $p_T$ requirement on $\bar b$ becomes $8-9$ GeV.  From 
Fig. ~\ref{fig1} at around $p_{T_{\rm min}}(\bar b) = 8-9$ GeV, 
the shaded CDF band gives 
\begin{equation}
\label{mc}
m_c \simeq  1.3 - 1.7\;{\rm GeV}\;,
\end{equation}
with the central value at about 1.45 GeV.
Since the error of the ratio in Eq. (\ref{bcfrag}) is large, the range of 
$m_c$ obtained in Eq. (\ref{mc}) is also very wide.

\section{}
\begin{figure}[t]
\centering
\includegraphics[width=6in]{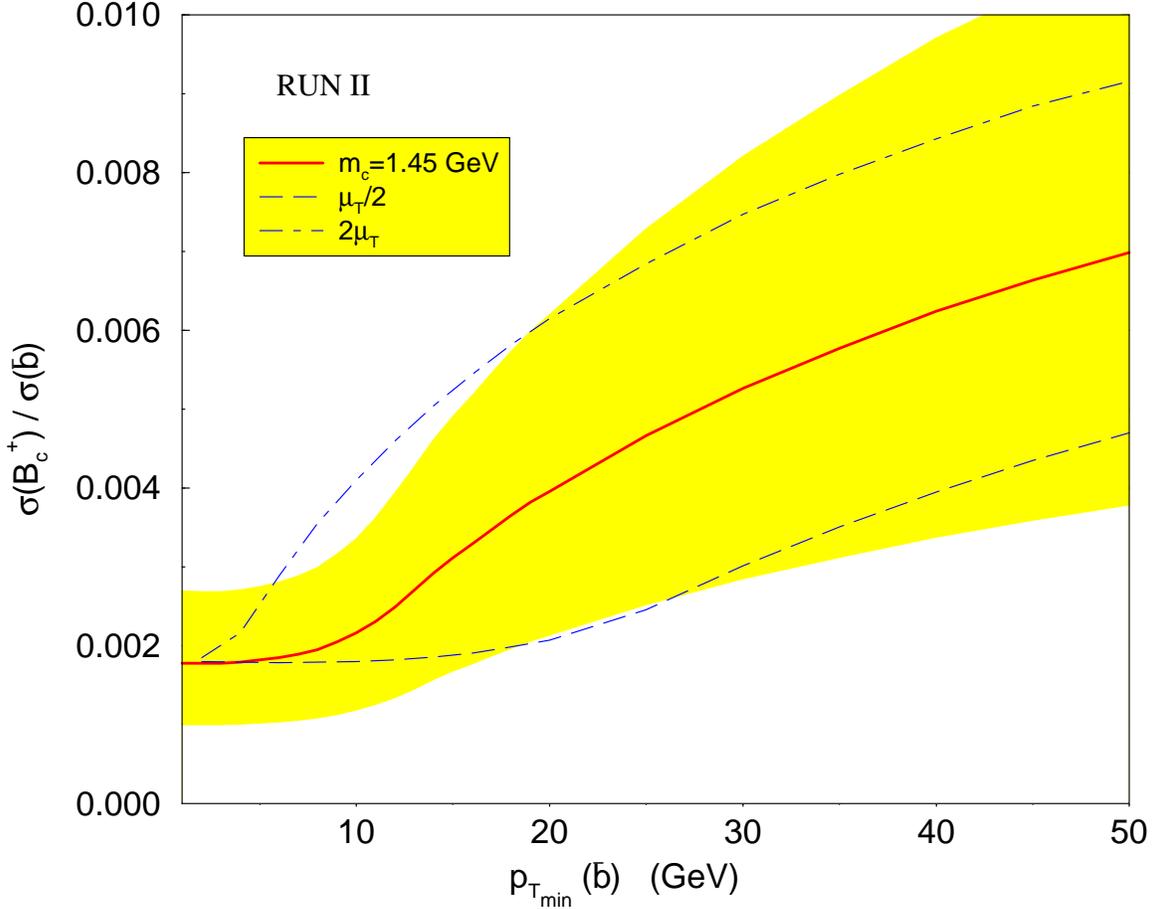}
\caption{\small 
The ratio of $\sigma(B_c^+)/\sigma(\bar b)$ versus $p_{T_{\rm min}}$ cut on
$\bar b$ calculated by fragmentation approach at Run II: $\sqrt{s}=2$
TeV.  The shaded region corresponds to $m_c \simeq 1.3 - 1.7$ GeV with
the solid line at $m_c = 1.45$ GeV. A rapidity cut of $|y|<1$ is imposed. 
The factorization scale $\mu=\mu_T \equiv \sqrt{p_{T_{\bar b}}^2 + m_b^2}$
for the solid line, $\mu= \mu_T/2$ for the dashed, and $\mu=2\mu_T$ for the
dot-dashed.
\label{fig2}}
\end{figure}

Run II at the Tevatron will be at $\sqrt{s}=2$ TeV with a nominal accumulated
luminosity of 2 fb$^{-1}$.  The prediction of $\sigma(B_c^+)/\sigma(\bar b)$
for the range of $m_c \simeq 1.3-1.7$ GeV obtained above in Eq. (\ref{mc})
is given in Fig. \ref{fig2} (shaded region) with the solid line for 
$m_c=1.45$ GeV.  It appears that the ratio predicted at 
$\sqrt{s}=2$ TeV is about the same as at $\sqrt{s}=1.8$ TeV.  

Finally, we also demonstrate the dependence of the ratio on the factorization
scale, which appears in the running $\alpha_s$, the parton distribution
functions, and the fragmentation functions.  In Fig. \ref{fig2}, the solid 
line is for the original choice of $\mu=\mu_T \equiv \sqrt{p_{T_{\bar b}}^2+ 
m_b^2}$, while the dashed is for $\mu= \mu_T/2$ and
the dot-dashed for $\mu= 2\mu_T$.  The dependence on the scale is 
smaller than the dependence on $m_c$.

To summarize we have obtained the ratio $\sigma(B_c^+)/\sigma(\bar b) =
\left( 2.08 \err{1.06}{0.95}\right )\times 10^{-3}$ from the CDF data in
Eq. (\ref{data}).  We have also verified that the prediction by the 
perturbative QCD fragmentation approach is consistent with the CDF data, 
with $m_c \simeq 1.3-1.7$ GeV and the central value at 1.45 GeV.  
The prediction of the ratio at Run II is very similar to that at Run I.

\section*{Acknowledgments}
We would like to thank Steve Mrenna for helpful discussions.
This research was supported in part by the U.S.~Department of Energy under
Grants No. DE-FG03-91ER40674 and by the Davis Institute for High Energy 
Physics.



\end{document}